\documentclass[aps,prd,preprint,superscriptaddress]{revtex4}

\usepackage{graphicx}
\usepackage{amsmath}
\usepackage{bm}
\usepackage{color}

\def\ee{\end{eqnarray}}

\def\=:{=\hspace{-.7em}\raisebox{1.1ex}{.}\hspace{.1em}\raisebox{-0.2ex}{.} }

\def\ee{\end{eqnarray}}

\def\=:{=\hspace{-.7em}\raisebox{1.1ex}{.}\hspace{.1em}\raisebox{-0.2ex}{.} }
\newcommand{\NF}{N_{\rm F}}
\newcommand{\NC}{N_{\rm C}}

\newcommand {\beq}{\begin{eqnarray}}
\newcommand {\eeq}{\end{eqnarray}}
\newcommand {\non}{\nonumber\\}

\newcommand {\1}[1]{\frac{1}{#1}}

\newcommand {\ph}{\varphi}

\newcommand {\del}{\partial}

\begin{document}


\title{Josephson vortices and  the Atiyah-Manton construction
}


\author{Muneto Nitta}

\affiliation{
Department of Physics, and Research and Education Center for Natural 
Sciences, Keio University, Hiyoshi 4-1-1, Yokohama, Kanagawa 223-8521, Japan\\
}


\date{\today}
\begin{abstract}
We show that sine-Gordon solitons appear in the low-energy effective 
theory of a domain wall in a $U(1)$ gauge theory with two charged 
complex scalar fields with masses, 
if we introduce the Josephson interaction term between the scalar fields.  
We identify these sine-Gordon solitons as vortices 
or ${\bf C}P^1$ sigma model instantons in the bulk, 
which are absorbed into the domain wall world-volume.
These vortices can be called Josephson vortices since 
they appear in Josephson junctions of two superconductors. 
This setup gives a physical realization of 
a lower dimensional analogue of Atiyah-Manton construction 
of Skyrmions from instanton holonomy.  

\end{abstract}
\pacs{}

\maketitle

\section{Introduction}
The Skyrme model was proposed to describe nucleons as solitons (Skyrmions) 
in  the pion effective field theory or the chiral Lagrangian \cite{Skyrme:1962vh}. 
Although the nucleons are now known as bound states of quarks, 
the idea of the Skyrme model is still attractive.
In fact, the Skyrme model is still valid as the low-energy description of QCD, 
for instance, in holographic QCD \cite{Sakai:2004cn, Hata:2007mb}.
One of the difficulties of the Skyrme model is that 
no Skyrme solutions are available because of the non-integrability of the equation of motion, 
though construction of approximate solutions was proposed \cite{Houghton:1997kg}. 
Among several proposals,  
Atiyah and Manton gave a particularly interesting proposal that Skyrmion solutions 
can be approximated by the holonomy of Yang-Mills instantons 
\cite{Atiyah:1989dq}.  
It has been applied, for instance, to calculate 
the force between two Skyrmions from the two-instanton holonomy 
\cite{Hosaka:1990dg}.
While the physical meaning of this ansatz was unclear for long time, 
a physical ``proof" of the Atiyah-Manton ansatz was  
presented some years ago \cite{Eto:2005cc}. 
One can consider a non-Abelian domain wall 
in a certain $U(2)$ gauge theory in $d=5+1$ dimensions 
\cite{Shifman:2003uh}, the low-energy effective theory of which 
is the chiral Lagrangian at the leading order in 
its $d=3+1$ dimensional world-volume. 
The next leading order contains the Skyrme term \cite{Eto:2005cc}, 
which implies that the domain wall world-volume theory 
is the Skyrme model admitting Skyrmions within it.
It was shown that these Skyrmions are nothing but 
Yang-Mills instantons in the bulk point of view. 
Since we perform the integration along the codimension of the wall 
to obtain the effective wall world-volume theory, 
it gives a physical explanation of the Atiyah-Manton ansatz.

On the other hand, a lower dimensional analogue of 
the Atiyah-Manton ansatz was also proposed \cite{Sutcliffe:1992ep,Stratopoulos:1992hq}.
It was proposed that the sine-Gordon soliton can be approximately 
constructed as the holonomy of a ${\bf C}P^1$ instanton in $d=2+0$ dimensions
or a lump in $d=2+1$ dimensions. 
Since exact solutions of the sine-Gordon solitons are available, 
the lower dimensional Atiyah-Manton ansatz can be checked analytically, 
unlike the original Atiyah-Manton construction. 
It may help us to understand better or to refine the original proposal 
by Atiyah and Manton. 

In this paper, we give a physical realization of the lower dimensional 
Atiyah-Manton construction. 
We consider the $U(1)$ gauge theory 
coupled with two charged complex scalar fields $\phi_1$ and $\phi_2$ 
with masses in $d=2+1$ dimensions,  
which reduces to the ${\bf C}P^1$ model in the strong gauge coupling limit. 
This model can be supersymmetric by properly adding bosonic and fermionic fields 
\cite{AlvarezGaume:1983ab}. 
This model is known to admit 
a domain wall solution \cite{Abraham:1992vb, Arai:2002xa}.
We add a deformation term $\phi^{1*} \phi^2$ in the original Lagrangian 
which breaks supersymmetry. 
This term is known as the Josephson term in 
the Josephson junction of two superconductors with two condensates 
$\phi^1$ and $\phi^2$. 
We show that this term induces the sine-Gordon potential 
in the effective theory of the $d=1+1$ dimensional domain wall world-volume. 
We find that the sine-Gordon soliton in the domain wall world-volume 
is nothing but an instanton or a lump in the ${\bf C}P^1$ model 
or a vortex in the gauge theory in the $d=2+1$ dimensional bulk. 
We call this object the Josephson vortex. 
This terminology is borrowed from the Josephson junction. 

Kinks inside a domain wall were also studied in supersymmetric gauge theories 
\cite{Ritz:2004mp, Eto:2005sw, Auzzi:2006ju, Bolognesi:2007bh}.
In particular, our work is closely related to a previous work \cite{Auzzi:2006ju}, 
in which an ${\cal N}=1$ supersymmetry preserving deformation 
term of ${\cal N}=2$ supersymmetry was considered in $d=3+1$. 
The domain wall is precisely the same as ours \cite{Abraham:1992vb, Arai:2002xa} 
without the deformation.  
In that model too, the effective theory of the domain wall 
is the sine-Gordon model, 
and the flux absorbed in the domain wall is a sine-Gordon soliton.  
However the crucial difference with ours is that 
the minimum flux inside the domain wall is half-quantized in 
their case, while it is unit quanta in our case.

In the limit that the domain wall is infinitely heavy,
our model is close to a Josephson junction of 
two superconductors of two condensations $\phi_1$ and $\phi_2$ 
sandwiching an insulator.  
Vortices in the bulk are absorbed into the insulator, 
becoming Josephson vortices or fluxons; 
see Ref.~\cite{Ustinov:1998} as a review. 
As in our case, dynamics of Josephson vortices 
can be described by the sine-Gordon equation. 
Josephson vortices also appear in high-$T_c$ superconductors 
with multi-layered structures \cite{Blatter:1994} and in 
two coupled Bose-Einstein condensates \cite{Kaurov:2005}.

In addition, 
a kink inside a domain wall appears in several systems in condensed matter physics: 
a Bloch line in a Bloch wall in magnetism \cite{Chen:1977}, chiral p-wave superconductors, 
and a Mermin-Ho vortex within a domain wall in 
$^3$He superfluid (see Fig.~16.9 of Ref.~\cite{Volovik2003}). 
Therefore, our method of a field theoretical approach 
may be applied to these condensed matter systems.

This paper is organized as follows. 
After our model is given in Sec.~\ref{sec:model}, 
we present the main results in Sec.~\ref{sec:SGkink}; 
we construct the domain wall effective theory by 
the moduli approximation of Manton \cite{Manton:1981mp} 
and find it to be the sine-Gordon model 
when we add the Josephson term in the original theory. 
We then construct sine-Gordon kinks, 
and show that they carry instanton (lump) charge in the bulk. 
Section \ref{sec:summary} is devoted to a summary 
and discussion. 
An application to the Atiyah-Manton construction is briefly discussed.

\section{The model \label{sec:model}}
We consider the $U(1)$ gauge theory 
coupled with two charged complex scalar fields $\phi^1(x)$ and $\phi^2(x)$ 
with masses and real scalar field $\Sigma(x)$ in $d=2+1$ dimensions. 
The Lagrangian which we consider is given by
\beq
&& {\cal L} = -\1{4 e^2} F_{\mu\nu}F^{\mu\nu} 
 + \1{e^2} (\partial_{\mu} \Sigma)^2 
 + |D_{\mu} \Phi|^2 - V\\
&& V = {e^2\over 2} (\Phi^\dagger\Phi -v^2)^2 
 + \Phi^\dagger(\Sigma {\bf 1}_2 - M)^2 \Phi 
 - \beta^2 \Phi^\dagger \sigma_x \Phi  
\eeq
where $e$ is the gauge coupling, 
complex scalar fields are written as $\Phi^T = (\phi^1,\phi^2)$, 
and the masses are given by $M={\rm diag.}(m_1,m_2)$ with $m_1>m_2$.

We refer to the last term in the potential 
\beq 
{\cal L}_J= 
 \beta^2 \Phi^\dagger \sigma_x \Phi 
= \beta^2 \phi^{1*} \phi^2 + {\rm c.c.}
\eeq 
as the ``Josephson" interaction term, 
because it appears in the Josephson junction of two superconductors 
with two condensations $\phi^1$ and $\phi^2$. 
In the limit $\beta=0$, the model enjoys 
${\cal N}=4$ supersymmetry (with eight supercharges) 
in $d=2+1$ with appropriately adding 
scalar fields $\tilde \Phi = (\tilde \phi^1,\tilde \phi^2)$ 
and fermion superpartners. 
In this case,  the Josephson term breaks supersymmetry explicitly. In this paper, supersymmetry is not essential apart from technical reasons \footnote{
We have added this term in order to stabilize the flux tube(vortex) 
absorbed into the domain wall in this theory.  
For $\beta=0$ the flux is diluted and disappears inside the domain wall. However, this term is not the unique term to stabilize it;
One could consider other renormalizable interactions. 
In this paper, we consider this term because it is the simplest among them. 
In fact, this term frequently appears in condensed matter physics 
such as Josephson junction of superconductors, high $T_c$ superconductors, 
and multi-component Bose-Einstein condensates. 
The other motivation to consider this term 
is an application to the Atiyah-Manton construction 
for which one-to-one correspondence between the flux tube 
outside the domain wall and the soliton inside the domain wall is required.
As discussed in Sec.~\ref{sec:extension}, an another term 
which preserves four supercharges was considered in Ref.~\cite{Auzzi:2006ju}. 
As we will see in the following section, one flux tube outside the domain wall 
corresponds to one soliton inside the domain wall in our case,
while one flux tube corresponds to two solitons inside the domain wall 
in the case of \cite{Auzzi:2006ju}. 
}.

For explicit calculation, 
we work in the strong gauge coupling limit $e^2 \to \infty$ 
in which the model reduces to the ${\bf C}P^1$ model 
with potential terms, 
but the results in this paper do not rely on this limit.
By rewriting $\Phi^T = (1,u)/\sqrt{1 + |u|^2}$ with 
complex projective coordinate $u$, 
the Lagrangian becomes 
\beq
&& {\cal L} =
 {\partial_{\mu} u^* \partial^{\mu} u - m^2 |u|^2 
  \over (1 + |u|^2)^2} + \beta^2 D_x , \quad
  D_x \equiv {u + u^*\over 1+|u|^2},\quad 
 \label{eq:CP1}
\eeq
with the mass $m \equiv m_1-m_2$. 
Here, $D_x$ is a moment map of the isometry generated by $\sigma_x$. 
With $\beta =0$, this model is known as the massive 
${\bf C}P^1$ model 
with the potential term of the Killing vector squared 
corresponding to the isometry generated by $\sigma_z$. 
It is a truncated version of 
a supersymmetric  sigma model with eight supercharges \cite{AlvarezGaume:1983ab}.
The potential of this model 
\beq
V= {m^2 |u|^2 
  \over (1 + |u|^2)^2} - {\beta^2  (u + u^*) \over 1+|u|^2}
\eeq
admits two discrete vacua $u=0$ and $u=\infty$ (for $\beta<m$).

Just for convenience, we can rewrite the Lagrangian 
in Eq.~(\ref{eq:CP1}) to another form. 
Introducing a three-vector of scalar fields by
${\bf n}(x) \equiv \Phi^\dagger \vec{\sigma} \Phi$ 
with the Pauli matrices $\vec{\sigma}$, 
the Lagrangian can be rewritten in the form of the $O(3)$ model:
\beq
 {\cal L} = \1{2} \partial_{\mu}{\bf n} \cdot \partial^{\mu}{\bf n} 
 - m^2(1-n_z^2) + \beta^2 n_x, 
 \quad {\bf n}^2 =1.
\eeq 
This model is known as the Heisenberg ferromagnet with 
anisotropy with two easy axes.

\section{Sine-Gordon Solitons from ${\bf C}P^1$ Instantons inside a Domain Wall
\label{sec:SGkink}}

\subsection{Domain wall solution}
For a while, we consider the case $\beta=0$, 
and we turn on it later. 
There are two discrete vacua $u=0$ and $u=\infty$.
Let us construct a domain wall perpendicular to the $x^1$-axis, 
interpolating these two vacua.
The Bogomol'nyi completion for the domain wall can be obtained as
\beq
E &=& \int d x^1 
\frac{
|\partial_{1} u \mp mu|^2 
\pm m (u^{\ast} \partial_1 u + u \partial_1 u^{\ast})}
{(1+|u|^{2})^{2}} \nonumber \\ 
 &\geq& |T_{\rm wall}| , \label{eq:BPS-bound-wall}
\eeq
where $\partial_i$ denotes 
the differentiation with respect to $x^i$. 
Here, $T_{\rm wall}$ is 
the topological charge which characterizes the wall:
\begin{eqnarray}
T_{\rm wall} = m \int  d x^1 \frac{ u^{\ast} \partial_1 u 
+ u \partial_1 u^{\ast}}{(1+|u|^{2})^{2}}   
= {m\over 2} \left[ {1-|u|^2 \over 1+|u|^2} \right]^{x^1=+\infty}_{x^1=-\infty}.
\end{eqnarray}
Among all configurations with a fixed boundary condition, 
that is, with a fixed topological charge $T_{\rm wall}$, 
the most stable configurations with 
the least energy saturate the inequality (\ref{eq:BPS-bound-wall}) 
and satisfy the 
Bogomol'nyi-Prasad-Sommerfield (BPS) equation  
\begin{equation}
\partial_{1} u \mp m u =0,  
\end{equation}
which is obtained by $|...|^2=0$ in  Eq.~(\ref{eq:BPS-bound-wall}). 
This BPS equation can be immediately solved as 
\cite{Abraham:1992vb, Arai:2002xa}
\beq
 u_{\rm dw} = e^{\pm m (x^1 - X) + i \ph} , \label{eq:wall}
\eeq
with the width $\Delta x^1 = 1/m$ and the tension 
\beq 
 |T_{\rm wall}|=m,
\eeq 
where $\pm$ denotes a domain wall and an anti-domain wall.
Here $X$ and $\ph$ are real constants called moduli parameters 
which are Nambu-Goldstone modes associated with 
broken translational and internal $U(1)$ symmetries, respectively.

\subsection{Low-energy effective theory on domain wall world-volume}
Next, let us construct the effective field theory of the domain wall 
($+$ signature in Eq.~(\ref{eq:wall})). 
According to Manton \cite{Manton:1981mp}, 
the effective theory on the domain wall can be obtained 
by promoting the moduli parameters to fields $X(x^i)$ and $\ph(x^i)$ 
on the domain wall world-volume $x^i$ ($i=0,2$)
and by performing the integration over the codimension $x\equiv x^1$:
\beq
 {\cal L}_{\rm dw.eff.} 
&=& \int_{-\infty}^{+\infty} dx 
 {e^{2mx} \over (1+e^{2mx})^2} [(\del_i X)^2 + (\del_i\ph)^2]\non
&=& \1{2m} [(\del_i X)^2 + (\del_i\ph)^2] - m,
\eeq
where the constant term recovers the domain wall tension. 
This is just a free field theory, or a nonlinear sigma model with 
the target space ${\bf R}\times S^1$.

Let us turn on the Josephson term ($\beta \neq 0$). 
We work in the parameter region $\beta \ll m$. We assume that the domain wall solution (\ref{eq:wall}) is not deformed.
The domain wall effective action is deformed by
\beq
 \Delta {\cal L} 
 &=& \beta^2 \int_{-\infty}^{+\infty} dx 
 {e^{mx+ i\ph} + e^{mx- i\ph} \over 1+e^{2mx}} \non
 &=& {2\over m} \int_{-\infty}^{+\infty} dx 
   {e^{mx} \over 1+e^{2mx}} 2 \cos \ph  
 = {\pi \beta^2 \over m} \cos \ph.
\eeq
Finally, we thus obtain the domain wall effective theory as
\beq
 {\cal L}_{\rm dw.eff.} 
 &=& \1{2m} [(\del_i X)^2 + (\del_i\ph)^2 + 2 \pi \beta^2 \cos \ph]\non
 &=& \1{2m} [(\del_i X)^2 + (\del_i\ph)^2 + \tilde \beta^2 \cos \ph], 
 \label{eq:SG}
\eeq
with $\tilde \beta^2 \equiv 2 \pi \beta^2$ apart from the constant term.
This is the sine-Gordon model with the additional field $X$.

\subsection{The sine-Gordon soliton inside the domain wall}
\begin{figure}[ht]
\begin{center}
\begin{tabular}{cc}
\includegraphics[width=0.6\linewidth,keepaspectratio]{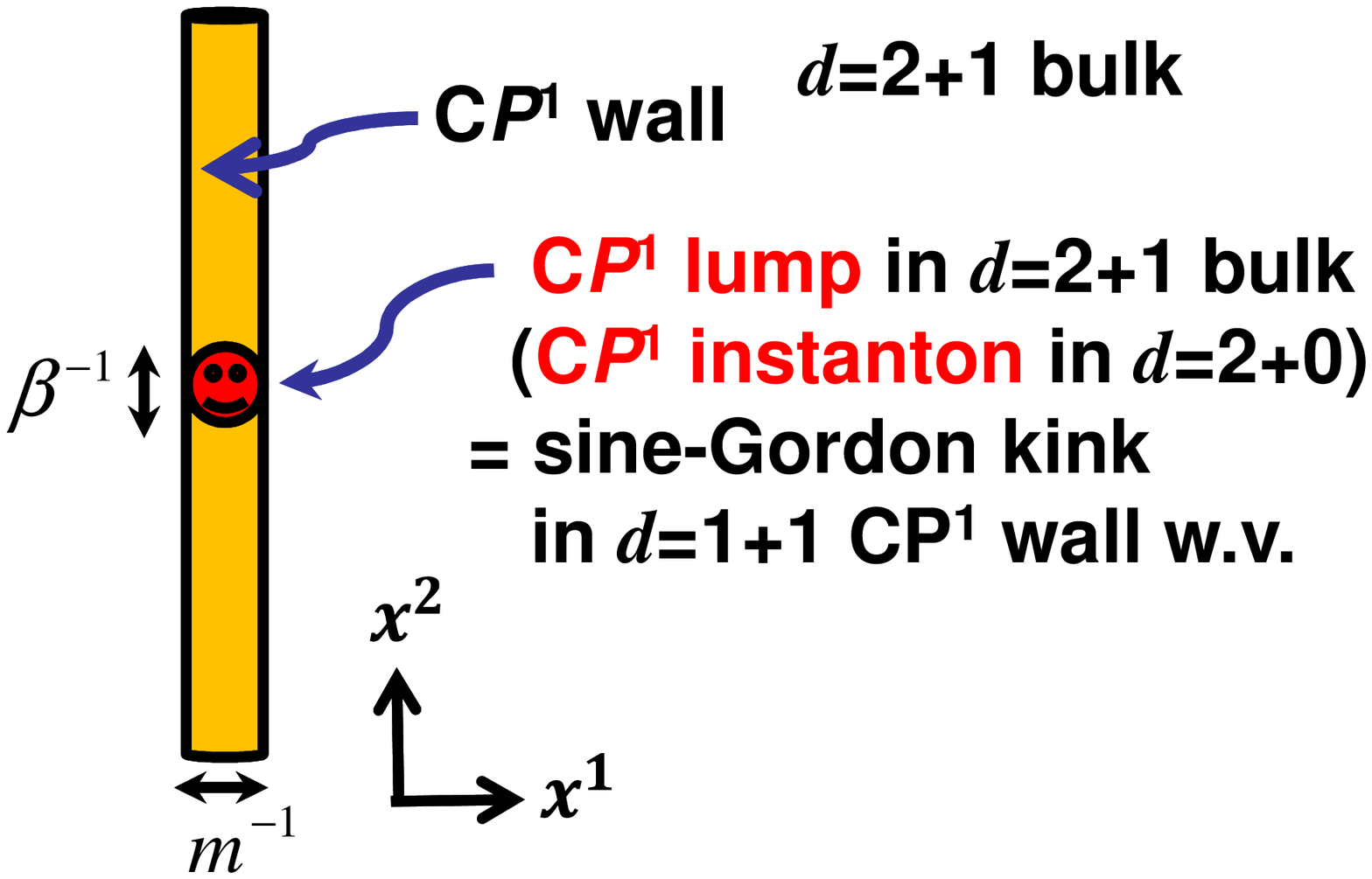}
&
\includegraphics[width=0.25\linewidth,keepaspectratio]{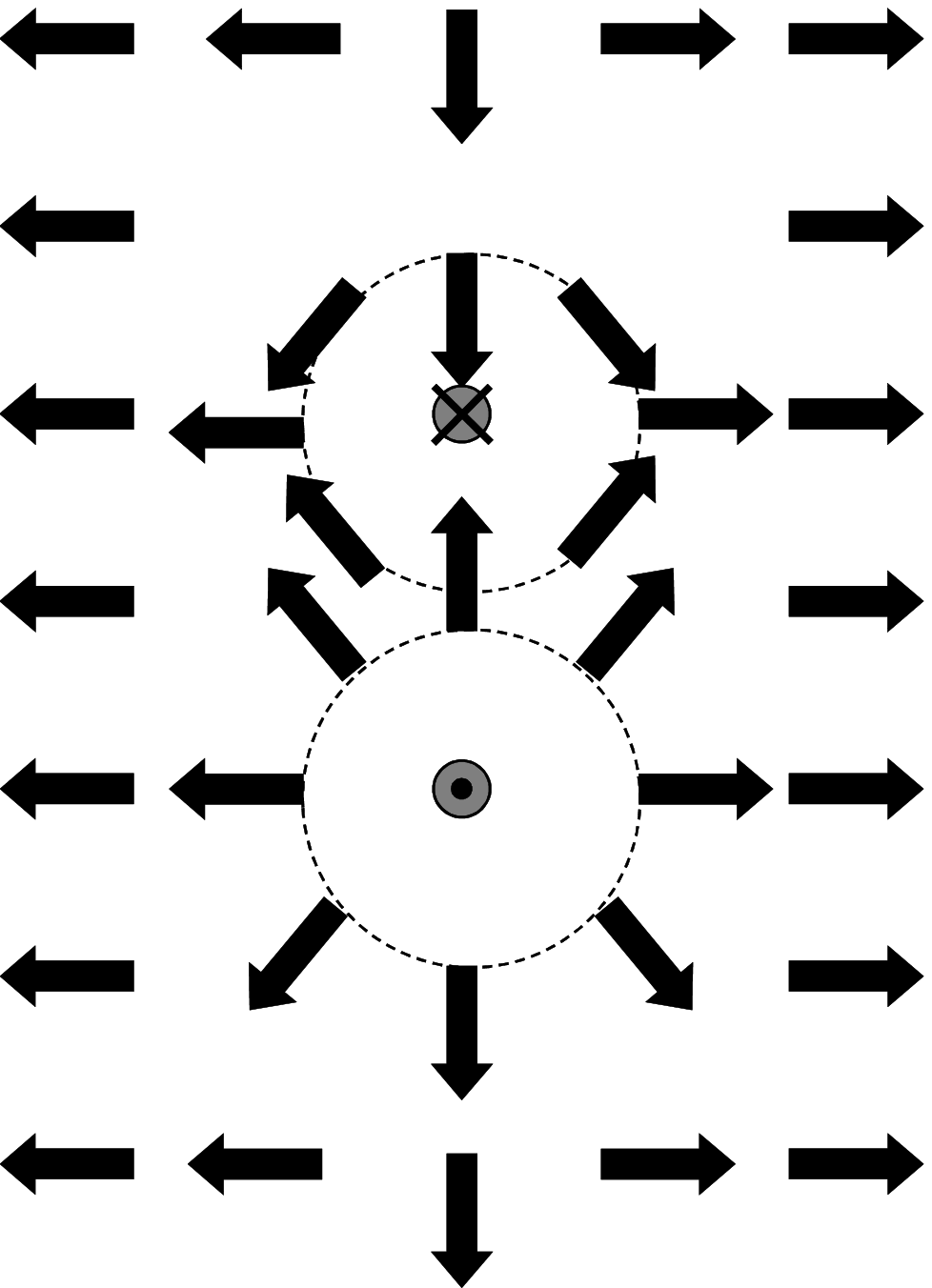}
\\
(a) & (b)
\end{tabular}
\caption{A sine-Gordon soliton in the domain wall describing 
the ${\bf C}P^1$ lump inside the domain wall. 
(a) Schematic configuration in the entire space.
(b) Points in the ${\bf C}P^1$ target space  are denoted by 
three-dimensional arrows. The north and south poles are denoted by 
the left and right arrows, respectively.
\label{fig:SG}}
\end{center}
\end{figure}

Next we construct a sine-Gordon kink in the domain wall effective theory 
and identify what it is in the bulk. 
The Bogomol'nyi completion for the energy density corresponding to 
the Lagrangian in Eq.~(\ref{eq:SG}) is obtained (for $X=0$) as 
\beq
 2m E &=& (\del_2 \ph)^2 + \tilde \beta^2 (\sin^2 {\ph\over 2} -1) \non
 &=& \left(\del_2\ph \pm \tilde \beta \sin {\ph\over 2}\right)^2 
  \mp 2 \tilde \beta \partial_2 \ph \sin {\ph\over 2} - \tilde \beta^2\non
 &\geq& 2m |t_{\rm SG}| - \tilde \beta^2
\eeq
with the topological charge density
\beq
t_{\rm SG} 
 \equiv { \tilde \beta \over m} \partial_2 \ph \sin {\ph\over 2}
 = - { 2 \tilde \beta \over m} \partial_2 \left( \cos {\ph\over 2}\right).
\eeq
The inequality is saturated by the BPS equation
\beq
 \del_2 \ph \pm \tilde \beta \sin {\ph\over 2} = 0.
\eeq
For instance, the one-kink solution can be given as
\beq
 \ph (x^2) = 4 \arctan \exp{{\tilde \beta \over 4} (x^2- Y)} 
 + {\pi \over 2}
\eeq
with the position $Y$ in the $x^2$-coordinate.  
The topological charge for this solution is 
\beq
 T_{\rm SG} = \int dx^2 t_{\rm SG} = {4\tilde \beta \over m}. 
\label{eq:SG-charge}
\eeq
The width of the sine-Gordon kink is $\Delta x^2 = 1/\tilde \beta$
so that we have a relation $\Delta x^1 /\Delta x^2 \sim m /\tilde \beta$. 
The total configuration is schematically drawn in  Fig.~\ref{fig:SG} (a). 
In Fig.~\ref{fig:SG} (b), we plot the spin texture of 
the ${\bf C}P^1$ target space 
for this configuration.

What is this solution in the $d=2+1$ dimensional  bulk  theory? 
We now show that this is a ${\bf C}P^1$ instanton (lump) in $d=2+1$. 
Let us calculate the topological lump (instanton) charge  by   ($a=1,2$)
\beq
 T_{\rm lump} 
&=& \int d^2x  {i (\partial_1 u^* \partial_2 u - \partial_2 u^* \partial_1 u )
\over (1+|u|^2)^2} \non
&=& \oint dx^a {-i (u^* \partial_a u - (\partial_a u^*) u )
             \over 2 (1+|u|^2)} \non
&=& \oint dx^a {|u|^2  \over 1+|u|^2} \partial_a \ph \non
&=& \int dx^2 \del_2 \ph |_{x^1=+\infty}
= [\ph]_{(x^1,x^2)=(+\infty,-\infty)}^{(x^1,x^2)=(+\infty,+\infty)}  \non
&=& 2 \pi k . \label{eq:lump-charge}
\eeq
Here we have used $\del_1 \ph =0$ at $x^2=\pm \infty$ in the third-to-last equality, 
and the $k$ winding  of the phase $\ph$ 
for $k$ sine-Gordon kinks in the last equality.  
This precisely shows the coincidence between 
the topological charges for $k$ lumps and $k$ sine-Gordon kinks.

Equivalently, this charge can be rewritten as the vortex charge 
\beq
&& T_{\rm vortex} = \int d^2 x F_{12} 
 = \oint dx^a A_a = T_{\rm lump}, \label{eq:vortex-charge}
\eeq
with the (auxiliary) $U(1)$ gauge field
\beq
A_{\mu} = {i\over 2} 
 (\Phi^\dagger \partial_{\mu} \Phi - (\partial_{\mu} \Phi^\dagger)  \Phi) 
 = {-i (u^* \partial_{\mu} u - (\partial_{\mu} u^*) u )
             \over 2 (1+|u|^2)} . 
\label{eq:gauge}
\eeq
If we work in finite gauge coupling $e$ instead of taking 
the infinite coupling limit, 
lumps are replaced with vortices with the charge in Eq.~(\ref{eq:vortex-charge}) 
counting the magnetic fluxes, 
where $A_{\mu}$ is dynamical gauge field which cannot be written as 
Eq.~(\ref{eq:gauge}).

Although the charges and the numbers 
of the sine-Gordon kinks in the wall and the lumps in the bulk 
coincide, the more detailed information, such as their shape, 
can be deformed. 
In fact,  the spin texture of the sine-Gordon kink in the domain wall 
shows that the lump is split into a pair 
of a vortex and an anti-vortex. 
Each of them has a half lump charge so that they 
are fractional lumps, that is, merons.

We have used the Bogomol'nyi completion to obtain the domain wall and 
sine-Gordon kinks. 
However, the composite state is not BPS anymore, 
because the Josephson term breaks supersymmetry. 
This implies the existence of the static interaction between 
the domain wall and the vortex in the bulk. 
Although both 
the sine-Gordon topological charge in Eq.~(\ref{eq:SG-charge}) 
and the lump charge in Eq.~(\ref{eq:lump-charge}) are 
proportional to the soliton number, 
their coefficients do not coincide. 
The former can be interpreted as the kink energy on the domain wall 
and the latter as the vortex energy in the bulk.
We thus find that the energies of the vortex are smaller inside the wall 
than in the bulk in the small $\beta$ regime ($\beta \ll m$) 
which we are working in. 
Therefore, we conclude that there exists the attraction between 
the vortex in the bulk and the domain wall 
and that the vortex is absorbed into the domain wall world-volume, 
becoming the stable Josephson vortex.

\subsection{Extension and related model}\label{sec:extension}
We can extend our model to $U(1)$ gauge theory 
coupled with $N$ (more than two) charged complex scalar fields $\phi^i(x)$ 
($i=1,\cdots,N$) 
with $M={\rm diag.} (m_1,\cdots,m_N)$ with $m_i > m_{i+1}$. 
A natural choice of Josephson terms may be introduced between two neighboring pairs \footnote{As another choice, Josephson couplings can be introduced 
for all pairs, in which case there is a molecule of vortices \cite{Nitta:2010yf}.}: 
\beq
 {\cal L}_J = \sum_{i=1}^{N-1} \beta^2_i \phi^{i*} \phi^{i+1} + {\rm c.c.}
\eeq
In the absence of  the Josephson terms, 
the model reduces to the massive ${\bf C}P^{N-1}$ model, 
admitting $N-1$ parallel domain walls   \cite{Gauntlett:2000ib, Isozumi:2004jc}.
With the Josephson terms, this describes arrays of $N$ Josephson junctions.  
Vortices (${\bf C}P^{N-1}$ instantons or lumps) in various components will be absorbed 
in each domain wall, which should be studied elsewhere. 

Finally let us make comments on the previous work \cite{Auzzi:2006ju},
where a nonlinear sigma model on 
the target space $(T^*) {\bf C}P^1$ is considered.
With the mass matrix 
$M=  {\rm diag.}(m,-m)$, 
the model admits a domain wall whose effective theory is a free theory,  
a sigma model on ${\bf C}^* = {\bf R} \times S^1$ in $d=1+1$ 
as ours. 
On the other hand, with the mass matrix 
$M= \left(\begin{matrix} m & - \beta /\sqrt 2 
\\ \beta/\sqrt 2 & - m \end{matrix}\right)$, 
the model admits a domain wall whose effective theory is the sine-Gordon model, 
namely ${\bf C}^*$ 
with a potential $V = - (\beta^2 \xi /m) {\rm cos}^2 \sigma$.
In this case, one sine-Gordon kink 
carries the half quantized flux of $U(1)$ gauge theory 
or the half lump (instanton) charge 
of the ${\bf C}P^1$ model.
Therefore, after one vortex in the bulk is absorbed into the domain wall, 
it splits into two sine-Gordon kinks in this case. 
On the other hand, in our model, the numbers of the sine-Gordon kinks 
in the domain wall and 
the instantons (lumps)  in the bulk correspond to each other one-to-one.

\section{Summary and Discussion \label{sec:summary} }

We have constructed a sine-Gordon kink in the domain wall world-volume 
in the $U(1)$ gauge theories 
coupled with two complex scalar fields $\phi^1$ and $\phi^2$ 
with the Josephson interaction term $\phi^{1*} \phi^2$
in $d=2+1$ dimensions. 
We have shown the sine-Gordon soliton in the  $d=1+1$ dimensional 
domain wall world-volume 
is nothing but an instanton or a lump in the ${\bf C}P^1$ model 
or a vortex in the gauge theory in the $d=2+1$ dimensional bulk. 
This provides a physical realization of the lower dimensional 
Atiyah-Manton construction. 

It was proposed in Ref.~\cite{Sutcliffe:1992ep} that 
a sine-Gordon kink $\ph$ (in $d=1+1$) is well approximated by 
a holonomy of ${\bf C}P^1$ instanton (in $d=2+0$):
\beq
&& (-1)^k \exp [i \ph(x)] 
 = \exp \left(\int_{-\infty}^{+\infty} A_1 (x^1,x^2) dx^1 \right) 
  \label{eq:ansatz} 
\eeq
with the instanton (lump) number $k$ of ${\bf C}P^1$ instantons 
with the auxiliary gauge field $A_1$ in Eq.~(\ref{eq:gauge}):
\beq
 \ph = k \pi + \int_{-\infty}^{+\infty} dx^1
{-i (u^* \partial_1 u - (\partial_1 u^*) u )
             \over 2 (1+|u|^2)}.
\eeq
From Fig.~\ref{fig:SG} (b), we expect that a better approximation 
will be given by a pair of a meron and an anti-meron rather than 
a cylindrically symmetric lump solution.  
This deformation may be achieved by considering caloron \cite{Eto:2004rz} 
(see also Ref.~\cite{Bruckmann:2007zh}), 
{\it i.e.}, a periodic lump solution on ${\bf R} \times S^1$ 
with taking the periodicity as the wall width $\Delta x^1 = 1/m$. 
Another improvement is 
replacing the ${\bf C}P^1$ model 
with lumps by a $U(1)$ gauge theory with two charged Higgs fields with semi-local vortices. 
This may give a better approximation because 
of an exponential rather than power law asymptotic behavior,
as discussed in Ref.~\cite{Sutcliffe:1992ep}.

If we extend the model to $N$ complex scalar fields,
reducing to the massive ${\bf C}P^{N-1}$ model 
in the strong gauge coupling limit, 
it admits $N-1$ parallel domain walls  \cite{Gauntlett:2000ib, Isozumi:2004jc}.
It remains as an interesting future work
how instantons are absorbed 
into each domain wall. 
Another interesting extension will be
non-Abelian $U(\NC)$ gauge theories with 
$\NF(>\NC)$ flavors in the fundamental representation 
($\NC \times \NF$ matrix of scalar fields).  
The model reduces to the massive Grassmannian 
$SU(\NF)/[SU(\NC)\times SU(\NF-\NC)\times U(1)]$ sigma model 
in the strong gauge coupling limit \cite{Arai:2003tc}.
 Appropriate extension of the Josephson terms is not known.
Without the Josephson terms, domain walls in this theory were studied in Ref.~\cite{Isozumi:2004jc}. 
The construction of the effective theory on general domain wall solutions can be found in Ref.~\cite{Eto:2006uw}.
The fate of Grassmannian lumps 
(or non-Abelian semi-local vortices \cite{Shifman:2006kd}) 
should be studied elsewhere. 

Yet another interesting extension will be to study 
what happens in the presence of domain wall junctions or networks \cite{Eto:2005cp,Eto:2006pg}, 
which are possible if we introduce complex masses $M$ 
for scalar fields, and domain walls stretched by vortices 
\cite{Isozumi:2004vg,Eto:2006pg}. 
The effective theories of the domain wall network and the vortices stretched between domain walls were constructed in Ref.~\cite{Eto:2006bb} and \cite{Eto:2008mf}, respectively.
How Josephson vortices are absorbed into these composite solitons remains for a future study. 

\section*{Acknowledgements}

This work is supported in part by 
Grant-in-Aid for Scientific Research (No. 23740198) 
and by the ``Topological Quantum Phenomena'' 
Grant-in-Aid for Scientific Research 
on Innovative Areas (No. 23103515)  
from the Ministry of Education, Culture, Sports, Science and Technology 
(MEXT) of Japan.


\newcommand{\J}[4]{{\sl #1} {\bf #2} (#3) #4}
\newcommand{\andJ}[3]{{\bf #1} (#2) #3}
\newcommand{\AP}{Ann.\ Phys.\ (N.Y.)}
\newcommand{\MPL}{Mod.\ Phys.\ Lett.}
\newcommand{\NP}{Nucl.\ Phys.}
\newcommand{\PL}{Phys.\ Lett.}
\newcommand{\PR}{ Phys.\ Rev.}
\newcommand{\PRL}{Phys.\ Rev.\ Lett.}
\newcommand{\PTP}{Prog.\ Theor.\ Phys.}
\newcommand{\hep}[1]{{\tt hep-th/{#1}}}

\end{document}